\documentclass[iop]{emulateapj}
\usepackage{amsmath}
\usepackage{amsfonts}
\usepackage{amssymb}
\usepackage{graphicx}
\usepackage{enumitem}
\begin{document}

\title{Starspots on WASP-85}
\author{T.~Mo\v{c}nik$^{1}$, B.~J.~M.~Clark$^{1}$, D.~R.~Anderson$^{1}$, C.~Hellier$^{1}$, and D.~J.~A.~Brown$^{2}$}
\affil{$^{1}$Astrophysics Group, Keele University, Staffordshire, ST5 5BG, UK\\
$^{2}$Department of Physics, University of Warwick, Coventry, CV4 7AL, UK}
\email{t.mocnik@keele.ac.uk}
\shorttitle{Starspots on WASP-85}
\shortauthors{Mo\v{c}nik et al.}

\begin{abstract}
By analysing K2 short-cadence observations we detect starspots on WASP-85A, the host star of the hot Jupiter WASP-85Ab. The detection of recurring starspot occultation events indicates that the planet's orbit is aligned with the star's rotational axis ($\lambda<14^{\circ}$) and suggests a stellar rotational period of 15.1\thinspace $\pm$\thinspace 0.6\thinspace d. The K2 lightcurve reveals a rotational modulation with a period of 13.6\thinspace $\pm$\thinspace 1.6\thinspace d, consistent with the period determined from starspots. There are no significant transit-timing variations and thus no evidence of any additional planet in the system. Given the pronounced rotational modulation we are only able to place an upper limit of 100 parts per million for any phase-curve modulations and the secondary eclipse.
\end{abstract}


\section{INTRODUCTION}

Since the failure of a second reaction wheel, the \textit{Kepler} telescope provides the community with K2 observations of fields along the ecliptic plane \citep{Howell14}, and is therefore observing some previously known WASP exoplanets. One result has been the discovery of two additional transiting planets in the WASP-47 system \citep{Becker15}.

The transiting hot Jupiter WASP-85Ab was discovered by \citet{Brown14} and was observed by K2 in Campaign 1. It has a 2.66-d orbit around its G5, \textit{V}\thinspace =\thinspace 11.2 host star. WASP-85A forms a close visual binary, with an angular separation of 1.5\thinspace arcsec, with a cooler and dimmer K0, \textit{V}\thinspace =\thinspace 11.9 WASP-85B. The WASP photometry showed that one of the two binary components exhibits a 14.6\thinspace $\pm$\thinspace 1.5\thinspace d rotational modulation, which indicates stellar activity \citep{Brown14}.

When a starspot is occulted by a transiting planet the lightcurve exhibits a ``bump'' of temporary brightening \citep{Silva03}. The detection of starspots can provide a measurement or a constraint on the stellar obliquity, i.e. the angle between the stellar rotation and the planet's orbital axis. For example, the recurrence of the same starspot in consecutive transits has revealed that Kepler-17 has a low obliquity \citep{Desert11}. The ability to detect the recurring starspot occultation in a particular system depends on the ratio between the stellar rotational and planet orbital periods, and the lifetime of starspots. Alternatively, if starspots are occulted at certain preferential phases of the transit, this may constrain the obliquity if the transit chord crosses the active stellar latitudes at the observed active phases, as in the case of HAT-P-11 \citep{Sanchis11}.

The majority of hot Jupiters around stars cooler than around 6250\thinspace K are found in aligned orbits, suggestive of a disc migration mechanism. Hotter stars, on the other hand, host hot Jupiters that are in many cases found in misaligned orbits \citep{Albrecht2012}, which cannot be accounted for solely by planet disc migration mechanism. Instead, it is believed that planet--planet scattering, Kozai mechanism and tidal dissipation play an additional role in shaping misaligned orbits of hot Jupiters (e.g. \citealt{Triaud10}).

In this paper, we present a detection of starspots in the K2 lightcurve of WASP-85, which we use to constrain the stellar obliquity. We also refine the system parameters and the stellar rotational period, and search for transit-timing variations, as could be caused by an additional body in the system.

\section{THE K2 OBSERVATIONS}

The K2 Campaign 1 observations cover an 82-d time-span between 2014 May 30 and 2014 August 20 and provide a total of 120\thinspace 660 short-cadence images. We retrieved the 1-min short-cadence target pixel file for WASP-85 via the Minkulski Archive for Space Telescopes (MAST).

Given the degraded pointing accuracy of K2, the photometric precision is reduced because the position of an observed object drifts within the photometric aperture, and because of the non-uniform pixel sensitivity of the on-board detectors. \citet{Vanderburg14} developed a self-flat-fielding (SFF) procedure with which they correct the long-cadence photometry to within a factor of two of the original \textit{Kepler} precision by correlating the measured flux and arclength of the drift. Our K2 data-reduction procedure for short-cadence observations is similar to the procedure presented by \citet{Vanderburg14}.

\begin{figure*}
\includegraphics[width=\textwidth]{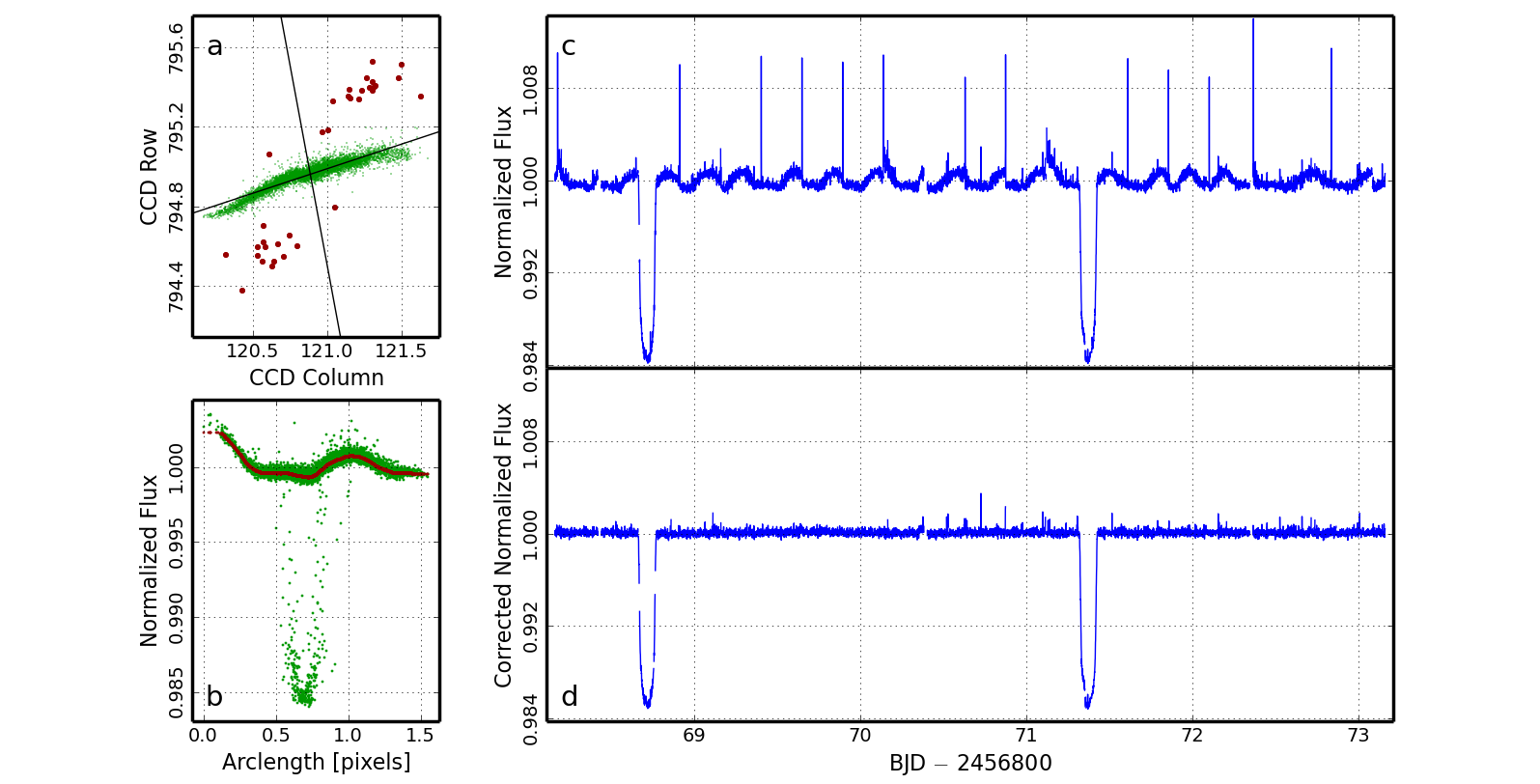}
\caption{SFF calibration procedure. All the panels correspond to a time interval of 5\thinspace d. 1a: Position of the center of the object given as CCD row vs. CCD column. Spatial outliers are marked with red and flagged as thruster events. 1b: Measured normalized flux vs. arclength of the spacecraft drift. The correlation is fitted with a Gaussian convolution (marked with red). 1c: Lightcurve before the SFF correction with the quality-flagged data points removed. 1d: Lightcurve after the SFF correction with the quality-flagged data points removed, including the thruster events which were identified during the SFF procedure. The positive flux excursions in the corrected lightcurve are the remaining cosmic rays that have avoided being identified and quality-flagged as cosmic rays in the downloaded K2 target pixel file.}
\end{figure*}

We first applied a standard data reduction procedure using PyRAF tools for \textit{Kepler} (PyKE) version 2.6.2 \citep{Still12}. Besides removing all the frames that were originally quality-flagged by K2, we manually adjusted quality flags and removed an additional 100 frames between 2014 June 1 18:04\thinspace UT and 19:41\thinspace UT, since they were obtained while the spacecraft was still in coarse pointing after the attitude tweak. Next, we used a large 100-pixels aperture mask to accommodate the diluted photometric signal from both binary components, WASP-85A and WASP-85B, and to reduce the photometric losses due to the relative drift between the object and the extraction aperture. Photometry of the individual binary components could not be performed because the angular separation (1.5\thinspace arcsec) is smaller than the \textit{Kepler} point spread function (full width half maximum of 6~arcsec).

After the data extraction and removal of low-frequency modulations, we performed our SFF procedure. The procedure is based on the PyKE tool \textsc{kepsff} with two main modifications to adjust and improve the efficiency for the short-cadence data. First, the 1-min short-cadence frames obtained during the thruster firing events are seen to have object positions that are outliers from the mean position drift on the detector. Data points with positional deviations of more than 6$\sigma$ were flagged as thruster firing events and deleted (see Figure~1a). This level was established by trial and error, as rejecting the thruster firing events while removing a minimal number of good data points. A total of 490 data points were quality-flagged as thruster firing events, which represents 0.4 per cent of the entire lightcurve.

\begin{figure}
\includegraphics[width=8.5cm]{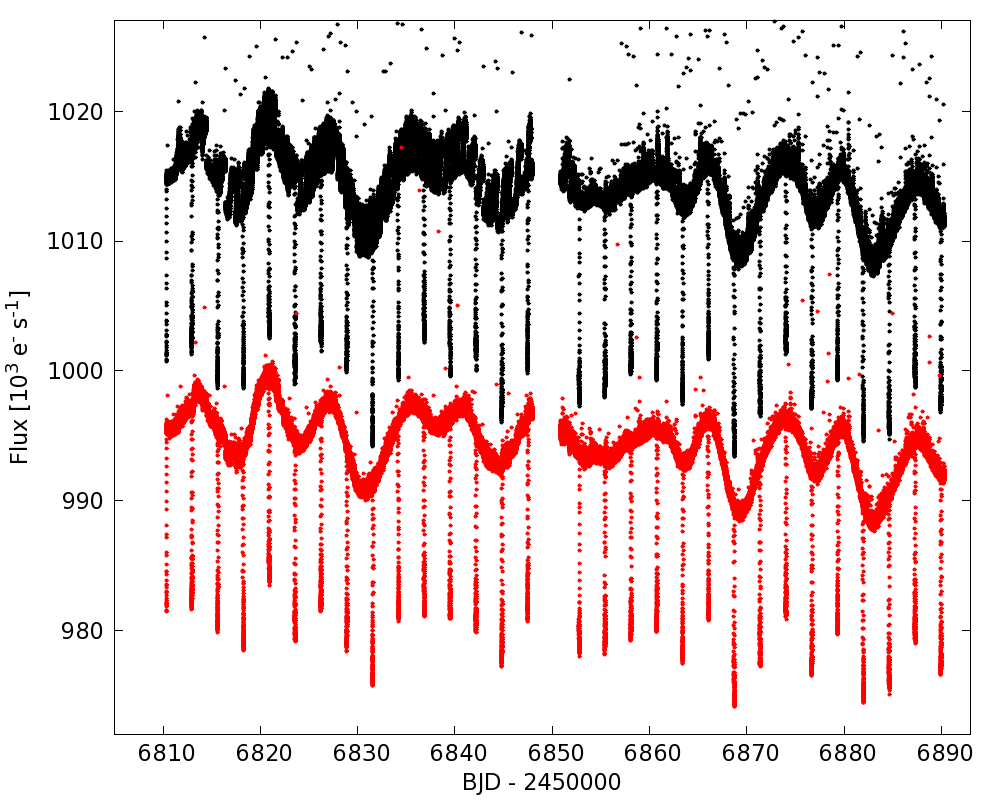}
\caption{K2 Campaign 1 short-cadence lightcurve of WASP-85 before (shown in black) and after the SFF correction (red). Note the presence of transits and starspot rotational modulation. The corrected lightcurve is shown with an offset of \mbox{$-$20\thinspace 000}\thinspace e\textsuperscript{-}\thinspace s\textsuperscript{-1} for clarity.}
\end{figure}

Second, instead of using the polynomial to fit the flux versus arclength of the positional drift of the spacecraft, we used convolution with a Gaussian kernel, which provided an improved fit of the correlation compared to the polynomial. For each data point the Gaussian convolution calculates the peak of the Gaussian flux distribution of a specified number of data points nearest in arclength to a particular data point. At this stage the deviating data points are typically of astrophysical origin and should be rejected to avoid any bias in the SFF calibration. Trial and error indicated that the flux-versus-drift arclength correlation was best fitted when using a Gaussian kernel with a width of 50 data points and a rejection threshold of 4$\sigma$ (see Figure~1b). After the Gaussian peaks have been calculated for all the data points the flux of every data point was replaced with a corresponding Gaussian peak. Figures~1c and 1d show an example lightcurve of a 5-d calibration time interval, before and after SFF, respectively. The time interval of 5\thinspace d was chosen as the best compromise between including a larger statistical sample of spacecraft drifts and a long-term variable drift behaviour.

With these two modifications we achieved nearly twice the precision improvement compared to the original \textsc{kepsff} PyKE tool for the short-cadence dataset of WASP-85. The median 1-min photometric precision for WASP-85 (\textit{K}$_{p}$\thinspace =\thinspace 10.2) was 778 parts per million (ppm) before SFF and 260\thinspace ppm after SFF, an improvement by a factor 3. For a comparison, the 1-min photometric precision of the original \textit{Kepler} mission for a \textit{K}$_{p}$\thinspace =\thinspace 10.5 star was 202\thinspace ppm \citep{Still13}.

After removal of all the quality-flagged data points, i.e. images affected by the cosmic rays, coarse pointing, spacecraft attitude tweak, thruster firing and reaction wheel de-saturation events, and a downlink data gap near the middle of Campaign 1, the total number of useful data points was reduced from 120\thinspace 660 to 108\thinspace 872. Figure~2 shows the entire short-cadence lightcurve of WASP-85 before and after applying the SFF correction. A total of 30 transits are visible. Also visible is the rotational modulation caused by the presence of starspots.

\section{ANALYSIS AND RESULTS}
\subsection{System parameters}

We obtained the system parameters using the Markov Chain Monte Carlo (MCMC) code presented in \citet{CollierCameron07} and further described in \citet{Pollacco08} and \citet{Anderson15}. In addition to the high-quality K2 photometry presented herein we used the HARPS radial-velocity measurements of WASP-85A from \citet{Brown14}. These measurements are not thought to have been significantly contaminated by WASP-85B. We corrected the K2 lightcurve for dilution by WASP-85B using a flux ratio of 0.50 (see \citet{Brown14}). Because the presence of starspot occultations may lead to inaccurate determination of system parameters \citep{Oshagh13} we removed the starspot occultation events from the measured lightcurve prior to the MCMC analysis. Limb darkening was accounted for using a four-parameter law, with coefficients calculated for \textit{Kepler} bandpass and interpolated from the tabulations of \citet{Sing10}.

To refine the ephemeris (but not the other parameters), we produced a separate MCMC analysis that also included the ground-based photometry presented in \citet{Brown14}, with four-parameter limb-darkening coefficients from \citet{Claret00,Claret04}, as appropriate for different bandpasses. This extended the photometric baseline from 80\thinspace d to 6.5\thinspace yr and reduced the uncertainty on the period by a factor of 4.

We present the system parameters in Table~1 and superimpose the corresponding transit model on the K2 lightcurve in Figure~3.

There is a discrepancy between the stellar effective temperature derived from the HARPS spectra ($T_{\rm eff} = 5685 \pm 65$\thinspace K; \citealt{Brown14}) and the value obtained from fitting the transits in our MCMC analysis ($T_{\rm eff} = 6112 \pm 27$\thinspace K). The same discrepancy, albeit smaller, is present in the discovery paper, which used only the ground-based transits ($T_{\rm eff} = 5910 \pm 54$\thinspace K from MCMC; \citealt{Brown14}). When fitting the photometry (either ground-based or K2) with a transit model generated using the spectroscopic $T_{\rm eff}$, large systematics are evident during ingress and egress (see upper residual panel in Figure~3), suggestive of a poor modelling of the limb darkening. The $T_{\rm eff}$ discrepancy could be explained in part if light from WASP-85B had contaminated the spectra of WASP-85A, leading to a lower apparent $T_{\rm eff}$ for the primary, though any contamination is expected to be at a low level.

\begin{figure}
\includegraphics[width=8.5cm]{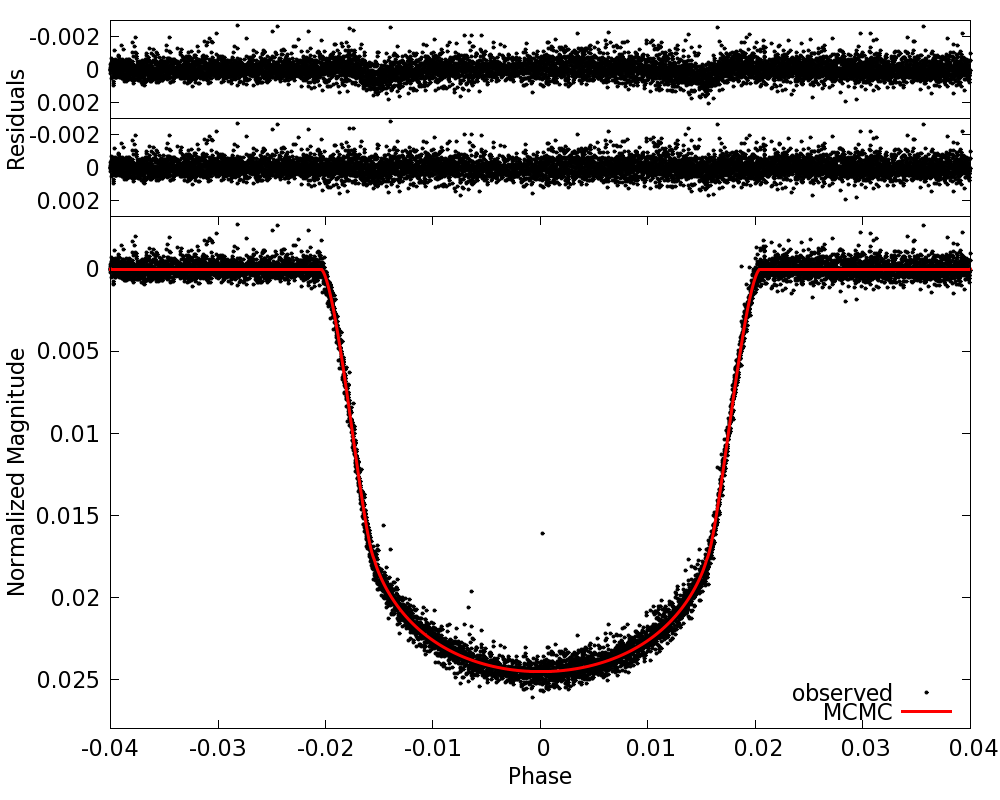}
\caption{Best-fit MCMC model and its residuals from the phase-folded K2 lightcurve near the transit. The lightcurve has been corrected for dilution by WASP-85B, using a flux ratio of 0.50. The lower residual panel corresponds to the effective temperature of 6112\thinspace K as fitted by our MCMC analysis. The upper residual panel serves as a comparison and corresponds to the spectroscopic effective temperature of 5685\thinspace K which results in large residuals near ingress and egress.}
\end{figure}

\begin{table*}
\centering
\begin{minipage}{11cm}
\caption{System parameters from MCMC for WASP-85Ab and its host star}
\begin{tabular}{@{}lccccc@{}}
\hline
Parameter&Symbol&Value&Unit\\
\hline
Transit epoch$^{a}$&\textit{t}$_{\rm 0}$&2456847.472856 $\pm$ 0.000014&BJD\\
Orbital period$^{a}$&\textit{P}&2.65567770 $\pm$ 0.00000044&d\\
Area ratio&$(R_{\rm p}/R_{\star})^{2}$&0.01870 $\pm$ 0.00002&...\\
Transit width&\textit{t}$_{\rm T}$&0.10816$^{+0.00006}_{-0.00002}$&d\\
Ingress and egress duration&\textit{t}$_{\rm 12}$, \textit{t}$_{\rm 34}$&0.0130377$^{+0.000008}_{-0.000016}$&d\\
Impact parameter&\textit{b}&0.047$^{+0.005}_{-0.018}$&...\\
Orbital inclination&\textit{i}&89.69$^{+0.11}_{-0.03}$&$^{\circ}$\\
Orbital eccentricity&\textit{e}&0 (adopted)&...\\
Orbital separation&\textit{a}&0.039 $\pm$ 0.001&AU\\
Stellar effective temperature$^{b}$&\textit{T}$_{\rm eff}$&6112 $\pm$ 27&K\\
Stellar mass&\textit{M}$_{\star}$&1.09 $\pm$ 0.08&M$_\odot$\\
Stellar radius&\textit{R}$_{\star}$&0.935 $\pm$ 0.023&R$_\odot$\\
Stellar density&$\rho_{\star}$&1.330 $\pm$ 0.007&$\rho_\odot$\\
Planet equilibrium temperature&\textit{T}$_{\rm p}$&1452 $\pm$ 6&K\\
Planet mass&\textit{M}$_{\rm p}$&1.265 $\pm$ 0.065&M$_{\rm Jup}$\\
Planet radius&\textit{R}$_{\rm p}$&1.24 $\pm$ 0.03&R$_{\rm Jup}$\\
Planet density&$\rho_{\rm p}$&0.66 $\pm$ 0.02&$\rho_{\rm Jup}$\\
\hline
\end{tabular}
\begin{itemize}[leftmargin=0.35cm]
\setlength\itemsep{0cm}
\item[$^{a}$]Epoch and period derived by fitting the photometric datasets from the K2 and all the available ground-based observations.
\item[$^{b}$]This value, obtained from a fit to the transit lightcurve, is discrepant with the spectroscopic value ($T_{\rm eff} = 5685 \pm 65$\thinspace K).
\end{itemize}
\end{minipage}
\end{table*}

\subsection{TTV and TDV}

Transit-timing variations (TTVs) and transit-duration variations (TDVs) can arise from dynamical interactions with additional planets in the system (e.g. \citealt{Becker15} and citations therein). Typical reported TTV amplitudes for perturbed transiting exoplanets are of the order of a few tens of minutes with periods of a few hundred days \citep{Mazeh13}. TDVs on the other hand are expected to be observed with smaller amplitudes and in phase with TTV. The first unequivocal exoplanet TDV detection was measured for system Kepler-88 and indicated a TDV amplitude of 5\thinspace min and a period of 630\thinspace d \citep{Nesvorny13}.

We measured the transit-timings and transit-durations for WASP-85Ab for each of the transits using the MCMC analysis similarly as described in Section~3.1, and show the deviations in Figure~4. The detected starspot occultations have again been removed from the lightcurve prior to the analysis as in Section~3.1. The error bars in Figure~4 do not include any contribution due to the remaining unremoved stellar activity features, which will be contributing to the observed scatter of the TTV and TDV data points. One data point has been excluded from TTV and TDV analysis due to the thruster firing event residual that took place at the beginning of ingress of the fifth transit near BJD~2456821.

\begin{figure}
\includegraphics[width=8.5cm]{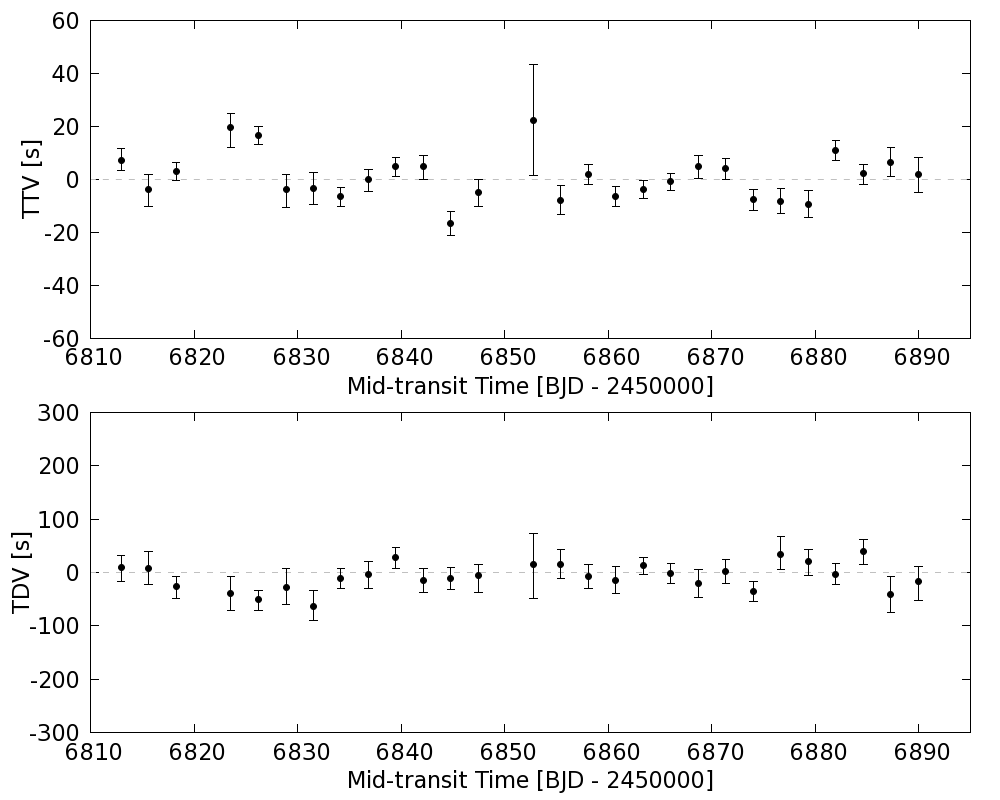}
\caption{TTV (upper panel) and TDV (lower panel) for WASP-85Ab. The measurements indicate a sinusoidal TTV variability with a small semi-amplitude of 7.8\thinspace s and a period 14.7\thinspace d, likely a signature of stellar activity rather than actual TTVs. The TDV measurements are consistent with the assumption of constant transit-duration. Note that the data point with the largest error bars corresponds to the transit during which a reaction wheel de-saturation took place near the ingress (see transit 17 in Figure~5).}
\end{figure}

The hypothesis of the TDV values being zero gives a $\chi^2$ of 31.8 for 28 degrees of freedom, which means we cannot reject it. The $\chi^2$ for TTV is 90.6 for 28 degrees of freedom which rejects the white noise distribution hypothesis and suggests a sinusoidal TTV signal with a semi-amplitude of 7.8\thinspace s and a period of 14.7\thinspace d. This TTV signal, if real, would to the best of our knowledge have the smallest detected TTV amplitude to date. On the other hand, it is known that stellar activity can induce TTV signals with semi-amplitudes of up to 200\thinspace s \citep{Oshagh13} and that the imperfect removal of some of the starspot occultation events, and the effect of the stellar rotational modulation, could have resulted in the TTV variability. The TTV period of 14.7\thinspace d is also compatible with the measured rotational period of $15.1 \pm 0.6$\thinspace d (see Section~3.4) and it is therefore more likely that the small measured TTV signal results from stellar activity.

The very small TTV amplitude and the absence of statistically significant TDV indicates that a second, non-transiting, massive planet is unlikely, unless at a much longer period.

\subsection{Starspot detection}

The short 1-min cadence of K2 observations and the restored photometric precision using the SFF enabled us to detect individual starspot occultations. We show in Figure~5 the transit lightcurves after subtracting the transit model, thus showing only the residuals. We first looked for starspots by calculating the $\chi^2$ value compared to a straight line. This was done by setting a box width equal to a phase width of 0.09, and sliding that box in time. The box width was determined by examining a few well defined starspot occultation events in the system. To estimate the threshold of significance we used the same box on out-of-transit data. Using this approach we were able to detect the 12 most prominent starspot occultation events above a $\chi^2_{\rm red}$ threshold of 4 (marked with asterisks in Figure~5). Below this threshold the residual lightcurve proved to be too unstable for this occultation detection technique.

To identify additional occultation events we asked four colleagues to mark definite and possible occultation events in the residual lightcurves near transits. The order of lightcurves was shuffled to avoid any bias towards spots that repeat in consecutive transits. All the events which were identified by at least two people as a possible starspot occultation event are listed in Table~2 and marked with red ellipses in Figure~5. We identified a total of 23 starspot occultations. The measured $\chi^2_{\rm red}$ of the individual occultation events are given in Table~2 and starspots are marked with different color intensities in Figure~5 as a rough indicator of their detectability.

\begin{figure*}
\includegraphics[width=\textwidth]{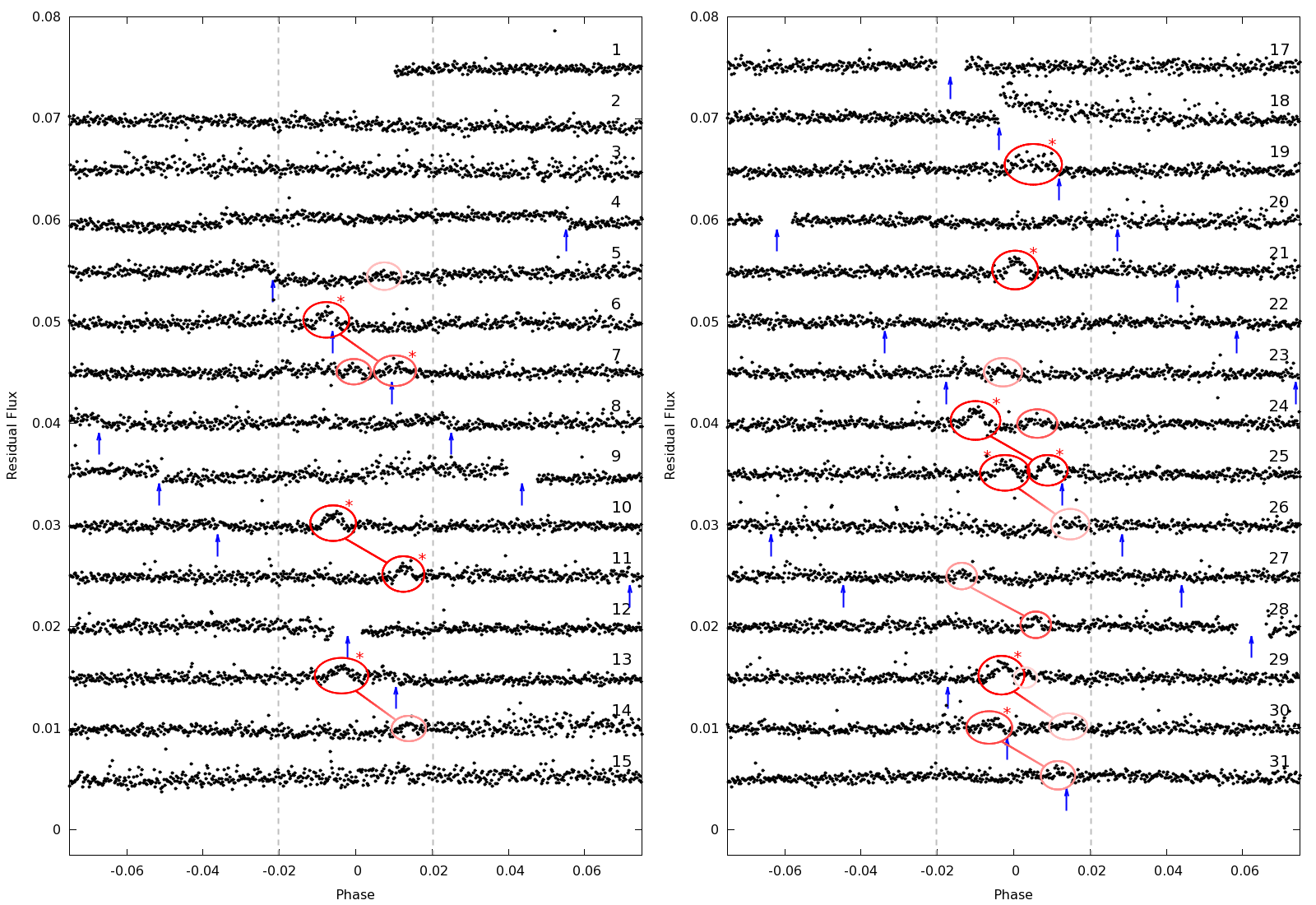}
\caption{Starspot occultations in the model-subtracted lightcurve of WASP-85Ab. Vertical dashed lines specify the extent of the transit. Red ellipses mark the 23 identified starspot occultation events. The color intensity denotes the detectability of events and asterisks correspond to events detected by our $\chi^2$ occultation detection technique. Potential starspot occultation pairs are marked with red lines. Blue arrows mark the positions of thruster firing events. Note that transit number 16 lies in the K2 data gap.}
\end{figure*}

Blue arrows are added in Figure~5 to mark the positions of thruster firing events which occasionally introduce artefacts in the lightcurve. Thruster firing artefacts are especially present in the first quarter of the observing campaign due to the somewhat inconsistent spacecraft drifts which might lead to a confusion with occultation events. Nevertheless, a typical potential thruster firing artefact only causes a small step-like deviation in the lightcurve and in the out-of-transit WASP-85 lightcurve never resembles a starspot-like bump. Therefore, it seems unlikely that any of the in-transit starspot-like event would be a thruster firing artefact, despite occasional overlaps.

\begin{table}
\centering
\caption{Starspot occultation events}
\begin{tabular}{@{}cccccc@{}}
\hline
Transit&$\chi^2_{\rm red}$&Phase&Stellar&Pair&Differential\\
number&&&longitude$^a$ [$^{\circ}$]&&longitude [$^{\circ}$]\\
\hline
5&1.57&0.0075&24.9&&\\
6&7.26&$-$0.0077&$-$25.6&6+7&60.7 $\pm$ 3.4\\
7&2.46&$-$0.0005&$-$1.7&&\\
7&4.91&0.0103&35.1&&\\
10&6.65&$-$0.0059&$-$19.4&10+11&63.7 $\pm$ 4.0\\
11&8.73&0.0125&44.3&&\\
13&5.25&$-$0.0037&$-$11.9&13+14&62.8 $\pm$ 4.2\\
14&1.65&0.0139&50.9&&\\
19&5.79&0.0051&16.7&&\\
21&6.21&0.0004&1.2&&\\
23&3.16&$-$0.0028&$-$9.0&&\\
24&8.13&$-$0.0100&$-$33.9&24+25&63.8 $\pm$ 3.8\\
24&3.06&0.0062&20.3&&\\
25&5.86&$-$0.0023&$-$7.5&25+26&63.3 $\pm$ 4.5\\
25&4.21&0.0089&29.9&&\\
26&1.82&0.0148&55.8&&\\
27&2.90&$-$0.0136&$-$49.4&27+28&68.2 $\pm$ 4.2\\
28&3.10&0.0058&18.8&&\\
29&5.63&$-$0.0028&$-$9.0&29+30&62.1 $\pm$ 4.3\\
29&2.31&0.0031&10.0&&\\
30&4.95&$-$0.0064&$-$20.9&30+31&61.3 $\pm$ 3.8\\
30&2.29&0.0143&53.1&&\\
31&1.81&0.0116&40.4&&\\
\hline
\end{tabular}
\begin{itemize}[leftmargin=0.35cm]
\setlength\itemsep{0cm}
\item[$^{a}$]Stellar longitude of zero corresponds to the meridian that runs through the center of the stellar disc.
\end{itemize}
\end{table}

\subsection{Are the starspots recurring?}

The presence of 23 starspots occultation events among 30 transits indicates that WASP-85A is magnetically active, that starspots occur uniformly along the transit chord, and that some starspot occultations are potentially occurring in pairs.

If the stellar rotational period is large compared to the orbital period of the planet and if the rotational and orbital axes are aligned, it is possible to see an occultation of the same starspot in consecutive transits. Figure~5 reveals eight such potential starspot occultation pairs at consistent phase shifts. For example, an occultation event at phase 0.013 during transit 11 was likely caused by the same starspot that was responsible for the occultation event at phase $-$0.006 in transit 10. All eight potential occultation pairs are marked with red lines in Figure~5 and their phase shifts and differential stellar longitudes are given in Table~2.

A detection of eight potential occultation pairs (16 spots) among a total of 23 starspot occultation events coupled with the small scatter among measured differential stellar longitudes indicates that occultation pairs were probably caused by occultations of the recurring starspots. It is therefore reasonable to suggest that planet's orbital and stellar rotational axis are aligned. Using the system parameter values from Section~3.1 we estimate that the obliquity angle has to be $<$14$^{\circ}$. From the mean differential stellar longitude of the starspot pairs of $63.2 \pm 2.3^{\circ}$ (see Table~2) and known orbital period we derive a stellar rotational period of $15.1 \pm 0.6$\thinspace d. This is in agreement with the rotational modulation period (see Section~3.5) and with the spectroscopically determined rotational period of 14.2\thinspace $\pm$\thinspace 4\thinspace d for WASP-85A \citep{Brown14}.

\subsection{Rotational modulation}

The K2 lightcurve shows a clear rotational modulation (see Figure~2). To estimate the period we used an autocorrelation function (ACF). For this task, we first removed the transits from the lightcurve and then applied 10$\sigma$ clipping. The rotational period is manifested as the highest ACF peak near 13.6\thinspace d and its multiples (see Figure~6). Peaks corresponding to half of this period are caused by starspot regions  on opposite sides of the star. The 80-d time-span of the K2 observations covers five cycles of the 13.6-d periodicity. Following the procedure demonstrated by \citet{McQuillan13} we determined the rotational period as a median of the intervals $\Delta\tau$ between the ACF peaks associated with the rotational period. We conservatively estimated the error as the FWHM of the highest ACF peak.

The resulting rotational period of $13.6 \pm 1.6$~d is in agreement with the spectroscopically determined rotational period for WASP-85A, 14.2\thinspace $\pm$\thinspace 4\thinspace d, if the star's rotational axis is orthogonal to the line of sight. But, the value is also in agreement with the estimated rotational period for WASP-85B, 11.7\thinspace $\pm$\thinspace 4\thinspace d \citep{Brown14}. Therefore, it is uncertain which of the two binary components is causing the observed rotational modulation. However, because WASP-85A is active (see Figure~5), it is reasonable to assume that the rotational modulation is caused by the rotation of the spotted surface of WASP-85A. Because rotational modulations depend strongly on starspots coming and going, which causes the phase shifts, the period derived from it is therefore less accurate and less reliable than the period determined via starspot occultations (see Section~3.4).

We used Spot Oscillation And Planet 2.0 (SOAP 2.0, \citet{Dumusque14}) to fit the starspot rotational modulation of WASP-85 and found that the three characteristic ``M''-shaped modulation features  can be well fitted with two main starspot regions on the opposite sides of the rotating stellar surface. However, the change in the lightcurve over time shows that the starspot regions are changing over the K2 observations.

\begin{figure}
\includegraphics[width=8.5cm]{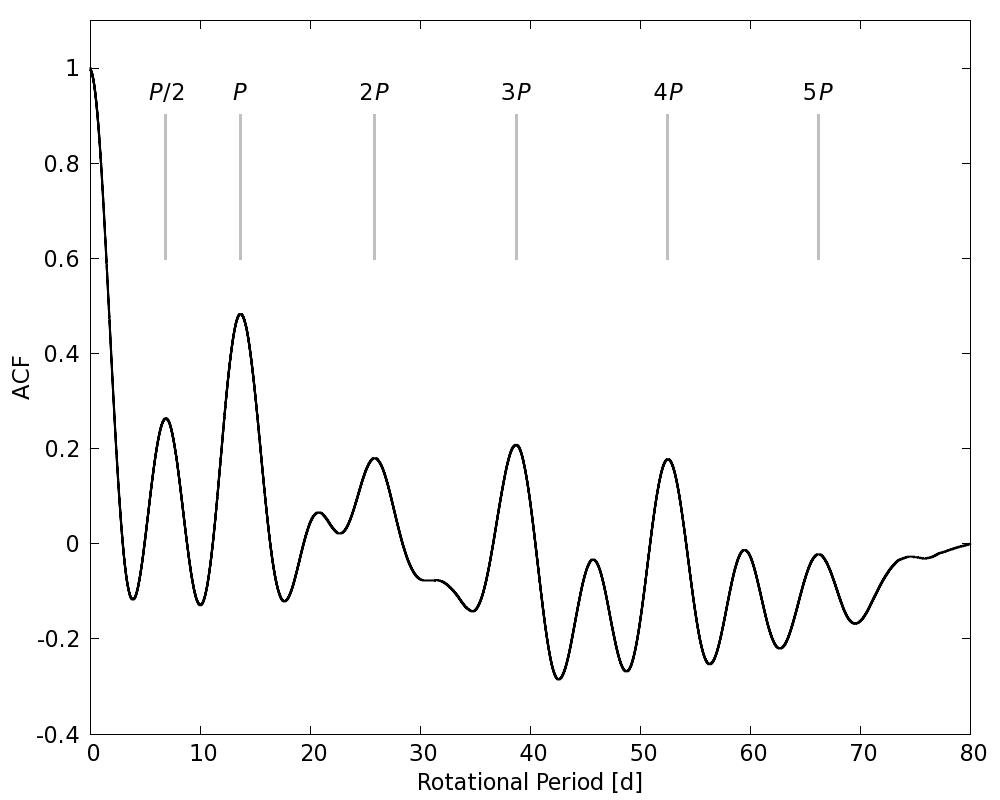}
\caption{Autocorrelation function of the K2 Campaign 1 short-cadence lightcurve of WASP-85. Marked with \textit{P} is the first and highest peak corresponding to 13.6\thinspace d rotational period. Additional four peaks are visible at multiples of this period. Peaks corresponding to half period are caused by another starspot region on the opposite side of the star.}
\end{figure}

\subsection{Phase-curve variations}

We folded the K2 lightcurve on the orbital period of the planet. In order to do this we first removed the rotational modulation using our two-region model of the starspots. However, since this model is only an approximation, the folded lightcurve is still dominated by the residuals of the starspot modulation, which do not average out owing to the limited length of the K2 observations (Figure~7).

We did not detect the secondary eclipse nor any other phase-curve variations caused by the orbiting exoplanet. Using the system parameters from Table~1 the expected semi-amplitudes of ellipsoidal, Doppler, and reflection variations are 3.7, 2.1, and (for an albedo of 0.1) 15.4\thinspace ppm, respectively \citep{Pfahl08}. We conservatively estimate an upper limit of 50\thinspace ppm for the semi-amplitude of variations over the planet's phase-curve. Figure~7 shows a comparison between the measured phase-curve of WASP-85 and the simulated reflection modulation with a semi-amplitude of 50\thinspace ppm. The inability to detect the reflection modulation and the secondary eclipse in the measured K2 phase-curve of WASP-85 indicates that the planet's visual geometric albedo is lower than $\sim$0.3.

\begin{figure}
\includegraphics[width=8.5cm]{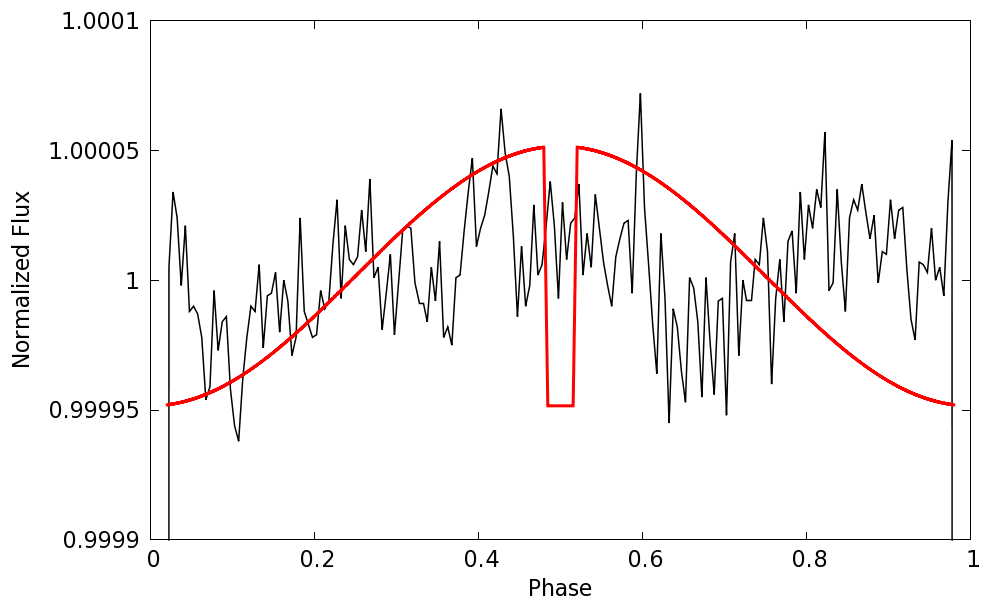}
\caption{Measured phase-curve of WASP-85, binned to 200 bins, (shown in black) and a simulated reflection phase-curve modulation with a semi-amplitude of 50\thinspace ppm (red). The inability to see the reflection modulation and secondary eclipse in the measured phase-curve of similar amplitude as shown in the simulated phase-curve suggests that the actual reflection semi-amplitude is lower than 50\thinspace ppm.}
\end{figure}

\section{CONCLUSION AND DISCUSSION}

We have modified the PyKE self-flat-fielding method to nearly double its performance in restoring the degraded short-cadence photometric precision due to the pointing drift of the K2 spacecraft. We find similar system parameters to the discovery paper \citet{Brown14}, measure a rotational modulation period of 13.6\thinspace d, and reconstruct the ``M''-shaped photometric modulations with two dynamic starspot regions on either side of a magnetically active star. The presence of starspot occultations indicates that the host star WASP-85A is active and suggests that the stellar rotation and the planet's orbital axis are aligned. The phase shifts of potential occultation pairs correspond to a stellar rotational period of $15.1 \pm 0.6$\thinspace d. The agreement between the stellar rotation and transit-timing variation period suggests that its small semi-amplitude of 8\thinspace s is a consequence of stellar activity rather than an inter-planet gravitational interaction. Measured transit-duration variability is consistent with white noise distribution around zero. We do not detect any phase-curve modulations nor the secondary eclipse of the planet.

Starspots provide one method of constraining the sky-projected obliquity of the host star, while the Rossiter-McLaughin (R--M) effect provides an independent estimate.

For example, in WASP-4 the R--M effect implies a projected alignment angle of $\lambda = -4^{\circ +43}_{\,\,\, -34}$ \citep{Triaud10}, while \citet{Sanchis11b} report that on two occasions the same starspot was occulted, from which they constrain the alignment to $\lambda = -1^{\circ +14}_{\,\,\, -12}$. Similarly in WASP-6 the R--M angle is $\lambda = 11^{\circ +14}_{\,\,\, -18}$ \citep{Gillon09}, and \citet{Tregloan15} suggest that the same starspot was occulted in two transits and report $\lambda = 7.2 \pm 3.7^{\circ}$. In WASP-19 the R--M angle is $\lambda = 4.6 \pm 5.2^{\circ}$ \citep{Hellier11}, while \citet{Tregloan13} saw the same starspot occulted on consecutive nights and report $\lambda = 1.0 \pm 1.2^{\circ}$.

HAT-P-11 is misaligned, with an R--M angle of $\lambda =
103^{\circ +26}_{\,\,\, -10}$ \citep{Winn10}. The misalignment is
confirmed by the analysis of starspots by \citet{Sanchis11}.
Similarly, for Kepler-63 \cite{Sanchis13} report $\lambda =
-110^{\circ +22}_{\,\,\, -14}$ from analysing both starspots and the R--M
effect.

There is currently no measurement of the R--M effect in our
system, WASP-85A, but we predict that it would confirm that the system
is aligned. Qatar-2 is a similar case, where starspots indicate alignment \citep{Mancini14} though this is not yet confirmed by an R--M measurement.

There are no reported cases where the R--M angle conflicts with the analysis of starspots, and thus the starspots are a reliable way of measuring alignment in systems too faint for R--M measurements. Examples are Kepler-30 (\textit{V}\thinspace =\thinspace 15.6) where occultations of starspots by both Kepler-30c and Kepler-30d indicate alignment \citep{Sanchis12}, and Kepler-17 (\textit{V}\thinspace =\thinspace 14.3) where starspots also indicate alignment \citep{Desert11}.

The benefit of ongoing K2 observations of known WASP planets is that the systems are generally bright enough to enable good measurements of the alignment angle, $\lambda$, the stellar rotational period, the stellar radius, and the star's projected rotational broadening, $v \sin i$. Given all of these we obtain, not only the projected alignment $\lambda$, but the actual alignment angle $\psi$ (see, e.g., \citealt{Winn05}). 

Adopting the rotational period of 15.1\thinspace $\pm$ 0.6\thinspace d, and combining with a stellar radius of 0.94\thinspace R$_{\odot}$ (Table~1) and a $v \sin i$ of 3.41 $\pm$ 0.89 km\thinspace s$^{-1}$ \citep{Brown14} we find that the star's rotational axis must be at $>$50$^{\circ}$ to the line of sight. This leads to the constraint that $\psi<41^{\circ}$.

The measured effective temperature and spin-orbit alignment of WASP-85 system are consistent with the current sample of predominantly aligned systems with stars below the characteristic effective temperature of 6250\thinspace K.

\acknowledgements{We would like to thank the anonymous referee for their comments which led to improving this paper. We gratefully acknowledge the financial support from the Science and Technology Facilities Council (STFC), under grants ST/J001384/1, ST/M001040/1 and ST/M50354X/1. This paper includes data collected by the K2 mission. Funding for the K2 mission is provided by the NASA Science Mission directorate. We thank Amaury Triaud for the use of RV data obtained with HARPS on the ESO 3.6-m telescope at La Silla (089.C-0151). This paper is based in part on observations made with the IO:O camera on the 2.0-m Liverpool Telescope (PL12B13). This work made use of PyKE \citep{Still12}, a software package for the reduction and analysis of \textit{Kepler} data. This open source software project is developed and distributed by the NASA Kepler Guest Observer Office.}

\bibliographystyle{apj}
\bibliography{bibliography}

\end{document}